\documentclass[aps,pre,preprint,unsortedaddress,letterpaper,showpacs,showkeys]{revtex4-1}

\usepackage{graphicx}
\usepackage{amsmath}
\usepackage{amssymb}
\usepackage{amsfonts}
\usepackage{dcolumn}
\usepackage{bm}
\usepackage{ulem}
\usepackage{color}

\begin{document}

\title{Unraveling the beautiful complexity of simple lattice model
  polymers and proteins using Wang-Landau sampling
\thanks{The research project is supported by the National Science
  Foundation under Grant No. DMR-0810223.}}




\author{Thomas W\"ust}
\email[]{thomas.wuest@wsl.ch}
\affiliation{Swiss Federal Research Institute WSL, Z\"urcherstrasse 111, CH-8903 Birmensdorf, Switzerland}

\author{Ying Wai Li}
\affiliation{Center for Simulational Physics, University of Georgia, Athens, Georgia 30602, U.S.A.}

\author{David P. Landau}
\affiliation{Center for Simulational Physics, University of Georgia, Athens, Georgia 30602, U.S.A.}

\date{\today}





\begin{abstract}

We describe a class of ``bare bones'' models of homopolymers which
undergo coil-globule collapse and proteins which fold into their
native states in free space or into denatured states when captured by
an attractive substrate as the temperature is lowered. We then show
how, with the use of a properly chosen trial move set, Wang-Landau
Monte Carlo sampling can be used to study the rough free energy
landscape and ground (native) states of these intriguingly simple
systems and thus elucidate their thermodynamic complexity.


\end{abstract}



\maketitle

\section{Introduction}

As physics orients itself increasingly towards soft condensed matter
and biology, the twin challenges of understanding how linear
homopolymers collapse and how proteins fold into their ground states
(``native states'' in the language of protein biochemistry) are being
attacked with ever increasing sophistication. Chain-like polymers are
often described by coarse-grained models that reduce the chemical
composition to simple ``monomers'' or ``beads'' connected in linear
fashion by strong chemical bonds \cite{polyreview}. As the
temperature is lowered, the non-bonded attraction between monomers
cause the polymers to collapse into densely entwined structures
\cite{polymer}. Proteins are polymer chains with complex side groups,
all composed of just a few kinds of atoms. In the same spirit,
proteins are often represented by simplified linear polymers with
\textbf{different} monomers in a particular sequence
\cite{protein_models}. Protein folding is a complex process which is
still poorly understood and which holds the key to unlocking a host of
biological and medical secrets. Both of these related types of
systems have the complication that the geometric constraints and
strong chemical bonds lead to entropic complexity in addition to
energetic considerations.

In principle the entire apparatus of statistical mechanics can be
brought to bear on the polymer collapse and protein folding problems,
but the determination of the thermodynamic properties and ground
states of each is still quite challenging. Methods of computer
simulation and improvements in computer performance have now reached a
level of maturity, however, that they can begin to illuminate the
polymer collapse or protein folding processes with good resolution.
(Of course, some of the clever techniques that have been developed for
magnetic systems, e.g. cluster flipping, do not translate directly to
use for the polymer/protein problem.) Hence, the focus of statistical
physics is rapidly shifting to the study of biological systems using
variations of some of the computational methods that have been well
developed for liquids and solids.

In order to permit study of finite temperature behavior, simple
polymer/protein models treat the atoms in the chains as classical
particles with strong chemical bonds and relatively weak interactions
between atoms that are not connected by chemical bonds. Nonetheless,
even simple bead-spring models with empirical bonded and non-bonded
potentials are still quite difficult to study at low temperature for
long chains \cite{polymer}. Hence, simple lattice models offer the
best hope of capturing polymer collapse
\cite{protein_models}. Proteins are even more difficult. Although
multiple microscopic, classical ``force fields'', i.e. interaction
Hamiltonians, have been developed, proteins are sufficiently complex
that these parametrized interaction fields are still of rather limited
value. In an attempt to make the problem more tractable,
Dill~\cite{dill} introduced a very simplified model which he believed
contained the essential features of the protein folding problem
without including all of the complexity. In Dill's approach, proteins
are modeled as lattice polymers which contain only two kinds of
monomers: hydrophobic ($H$) and polar ($P$). The goal was to provide
a model that was amenable to quantitative analysis and thus, it served
as a basis for - and inspired - computational studies of various
biophysical systems \cite{HPexamples}. However, it turns out that even
Dill's simplified protein model is very challenging for all but the
shortest sequences. Finding the ground states for an HP sequence is
proven to be an NP-hard problem \cite{NPhard}.

In contrast to molecular dynamics, which is generally restricted to
time scales of a few tens or hundreds of nano-seconds, clever Monte
Carlo algorithms, which do not follow a physical time scale, can be
used to explore behavior at much longer time scales
\cite{hansmann,landau_binder}. (Of course, for lattice models
molecular dynamics does not work at all.) The ``workhorse'' Monte
Carlo method in statistical physics, developed almost 60 years ago,
has been the Metropolis importance sampling algorithm
\cite{metropolis}; but more recently new, efficient algorithms have
begun to play a role in allowing simulation to achieve the resolution
which is needed to accurately locate and characterize ``interesting''
thermodynamic behavior
\cite{berg,polyreview,ferrenberg,janke_3,landau_binder}. Traditional
methods show quite long time correlations and require excessively long
runs to obtain precise results. As we shall see, for certain lengths
homopolymers may have complicated behavior as the temperature is
lowered. These difficulties also hold true for protein models that, as
expected, have complex free energy landscapes with energy barriers
between states of low free energy \cite{energylandscape}.
Consequently, ``standard'' Monte Carlo methods become easily
``trapped'' in metastable states from which they have difficulty
escaping.

Here we review a relatively new, general, efficient Monte Carlo
algorithm (known as ``Wang-Landau sampling" \cite{WL_PRL}) that offers
substantial advantages over other approaches \cite{rampf}. We will
then describe the application of this algorithm to the study of
lattice polymers of different chain length and then to the HP model
for several ``characteristic'' lattice protein sequences. One step
further in complexity, we will also present our work on the problem of
protein adsorption where we investigate HP model proteins interacting
with an attractive surface.

There have been a substantial number of other studies performed with
different simulational methods, and we shall try to place our results
in perspective.

\section{The lattice homopolymer and HP protein models}

Our lattice homopolymer model consists of a sequence of identical
monomers on sites of a periodic lattice, and these monomers are
connected together by unbreakable nearest neighbor bonds. The weak
inter\-actions between mono\-mers which are chemically separated but
physically nearby are modeled by an attractive nearest neighbor
potential; otherwise they do not interact. The HP model is a simple,
prototypical lattice protein model consisting of only two types of
monomers, hydrophobic (H) and polar (P), in a sequence chosen so as to
mimic a real protein~\cite{dill}. The chemical bonds between monomers
cannot break, and \textbf{non-bonded} interactions are restricted to
an attractive coupling, $\varepsilon_{HH}$, between hydrophobic
monomers that occupy nearest-neighbor sites. Although the model is
obviously over-simplified, it is intended to capture the hydrophobic
character, which is considered as the ``driving force'' for protein
folding \cite{hydrophobicity}. Moreover, in spite of the simplicity of
the model, for long chain lengths it is often very challenging to
determine the lowest possible energy (native state) for the folded
chain.

In the presence of an attractive surface, the homopolymer and protein
models can be easily extended to include the interaction between the
mono\-mers and the surface. (Here, we shall confine ourselves to the
protein case. For an overview of lattice homopolymers interacting
with surfaces, see Ref.~\cite{bachmann_janke_2005}.) In the simplest
case for proteins the surface attracts both types of monomers with
strength $\varepsilon_s$ and is introduced to the 3-dimensional
lattice as an $xy$-plane placed at $z = 0$. A second, non-attractive
wall is placed at $z = h_w$ to keep the HP chain from diffusing far
away from the bottom surface, but $h_w$ is still sufficiently large that the
protein can still be fully extended while in contact with the
surface. Periodic boundaries are used for the $x$ and $y$
directions. The Hamiltonian of the model is then represented as
\cite{bachmann_janke_2006}:
\begin{equation}
{\cal H} = -\varepsilon_{HH} n_{HH} - \varepsilon_s n_s,
\label{eq:hamiltonian}
\end{equation}
$n_{HH}$ being the number of H-H interacting pairs and $n_s$ being the
number of monomers adjacent to the bottom surface.

\section{The Wang-Landau sampling algorithm}

The shortcomings of traditional Monte Carlo algorithms led to the
development of a different, iterative approach to Monte Carlo sampling
which executes a random walk in energy space rather than in
probability space. This algorithm is now widely referred to as
``Wang-Landau sampling'' and has already been described in detail for
interacting magnetic spin systems \cite{WL_AJP}. The same basic
approach applies to lattice polymers/proteins, but suitable trial
moves must be implemented for the method to work efficiently
\cite{wuest}. Unlike the Metropolis Monte Carlo method, Wang-Landau
sampling does not depend upon the system temperature and focuses
instead on the density of states ${g(E)}$. If we perform a random
walk in energy space by moving monomers, and the probability to visit
a given energy level $E$ is proportional to the reciprocal of $g(E)$,
then a flat histogram of energies that have been visited is
generated. This is done iteratively by modifying the estimated
${g(E)}$ to produce a ``flat'' histogram of energies that have been
visited (over the allowed range of energy) while making ${g(E)}$
converge to the true value. During each succeeding step of the random
walk, $g(E)$ is modified and the updated value is used to continue the
random walk in energy space.

Initially, $g(E)$ can be set to $g(E)=1$ for all possible energies
$E$. Then a random walk in energy space is begun. If $E_1$ and $E_2$
are energies before and after a monomer is moved, the transition
probability from energy level $E_1$ to $E_2$ is
\begin{equation}
p(E_1 \rightarrow E_2)= \min\left({\frac {g(E_1)} {g(E_2)}}, 1\right). \label{eqn:p}
\end{equation}
Each time an energy level $E$ is visited, we multiply the existing
value of $g(E)$ by a ``modification factor'' $f>1$. (As a practical
matter it is preferable to work with the logarithm, i.e. $\ln(g(E))
\rightarrow \ln(g(E))+\ln(f)$, so that all possible values of $g(E)$
will fit into double precision numbers.) If a trial move is rejected,
the system stays at the same energy and the existing density of states
is multiplied by the modification factor. While multiple choices are
possible, a reasonable value of the initial modification factor is
$f=f_0=e^1\simeq2.71828\ldots$ During the random walk, we also
accumulate a histogram $H(E)$, i.e. the number of visits at each
energy level $E$, in energy space. When the histogram is sufficiently
``flat" over the entire energy range of the random walk, $g(E)$
converges to the true value with an accuracy proportional to that
modification factor $\ln(f)$. We then reduce the modification factor
using a function like $f_1=\sqrt{f_0}$, reset the histogram, and begin
the next iteration of the random walk. This process continues until
the histogram is ``flat" again and we then reduce the modification
factor $f_{i+1}=\sqrt{f_i}$ and restart. The process is halted when
the modification factor is smaller than some predefined value. (We
determined that $f_{{final}}=\exp(10^{-8})\simeq 1.00000001$ was a
good choice for the models considered here.) Note that the phrase
``flat histogram'' refers to a histogram $H(E)$ for which all entries
are not less than $x\%$ of the average histogram $\langle H(E)\rangle
$, where $x\%$ depends upon the desired accuracy of the density of
states.

During the early iterations the algorithm does not satisfy the
detailed balance condition, since $g(E)$ is constantly modified;
however, after many iterations, $g(E)$ converges to the true value as
the modification factor approaches $1$. If $p(E_1 \rightarrow E_2)$ is
the transition probability from the energy level $E_1$ to level $E_2$,
from equation (\ref{eqn:p}), the ratio of the transition probabilities
from $E_1$ to $E_2$ and from $E_2$ to $E_1$ can be calculated very
easily as
\begin{equation}
{\frac {p(E_1 \rightarrow E_2)} {p(E_2 \rightarrow E_1)}} = {\frac {g(E_1)} {g(E_2)}}
\end{equation}
where $g(E)$ is the density of states. Thus, the random walk algorithm
satisfies detailed balance. The proof of convergence, i.e. finding of 
the correct $g(E)$, of the Wang-Landau algorithm has been given rather 
in view of an optimization procedure than by means of Markov chain 
Monte Carlo arguments\cite{WL_proof}. In this respect the condition of 
detailed balance does not bear much importance (note that thermodynamic 
quantities are obtained only indirectly from $g(E)$). Nonetheless, for 
certain trial move sets, it is important to correct Eq.~(2) for unequal 
move ratios in order to fulfill detailed balance and to avoid a possible 
bias in the final estimate of $g(E)$\cite{wuest}.

\subsection{The importance of trial move sets}

\begin{figure}
\centering
\includegraphics[width=0.9\columnwidth]{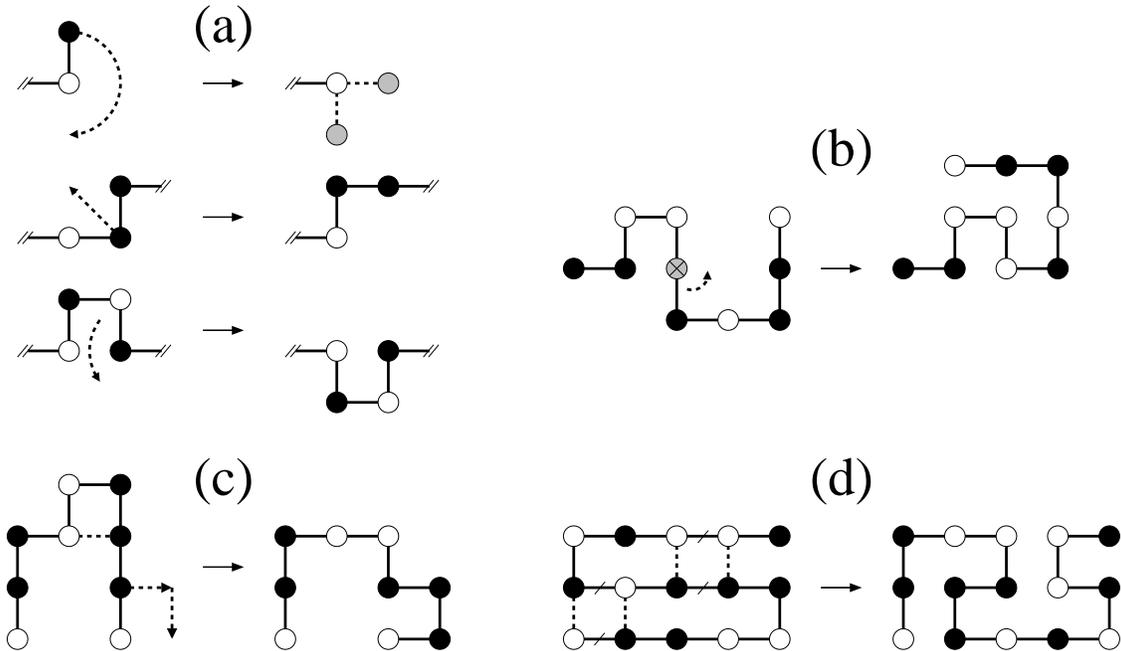}
\caption{\label{mcmoves}Examples of trial moves typically applied in
  Monte Carlo simulations of lattice polymers and proteins (here shown
  for a 2D sc lattice only). (a) End-bond flip, kink jump and
  crankshaft move; (b) pivot move; (c) pull move; (d) bond-rebridging
  move. These types of trial moves can be adopted to systems of any
  dimension $> 2$.}
\end{figure}

Simple trial moves that allow single monomer, or only a few
neighboring monomers, to move, (e.g. end-bond flips, kink jumps or
crankshaft moves, as shown in Fig. \ref{mcmoves}) have been developed
and used for many years for the study of geometric properties of
lattice polymers. These simple moves by themselves do not produce an
ergodic process, although the inclusion of ``pivot'' moves restores
ergodicity \cite{sokal}. This does not solve all the problems,
however, since the acceptance rate of ``pivot'' moves becomes
extremely low for compact polymers. For this reason, additional trial
moves need to be implemented, regardless of the Monte Carlo algorithm
used.

We have adopted two types of trial moves in our Wang-Landau sampling
simulations and determined that their use proves to be extremely
effective at accelerating convergence \cite{wuest}. The first kind of
trial move, ``pull moves'', effectively threads all of, or a portion
of, the polymer along the existing, entangled polymer structure
\cite{lesh}. A second kind of trial move, ``bond-rebridging'', does
not actually move any monomers but cuts bonds on opposite sides of an
elementary plaquette and redraws them on the other two sides to change
the polymer conformation dramatically in a single trial move
\cite{deutsch}. Since the monomers do not move, bond-rebridging alone
is not ergodic, but we shall combine pull moves with bond-rebridging
and ergodicity is no longer a problem. Fig.~\ref{mcmoves} shows
examples of ``traditional'' trial moves as well as of the ones applied
in our Monte Carlo simulations (pull and bond-rebridging moves).

\section{Three dimensional lattice homopolymers}

\begin{figure}
\centering
\includegraphics[width=0.85\columnwidth]{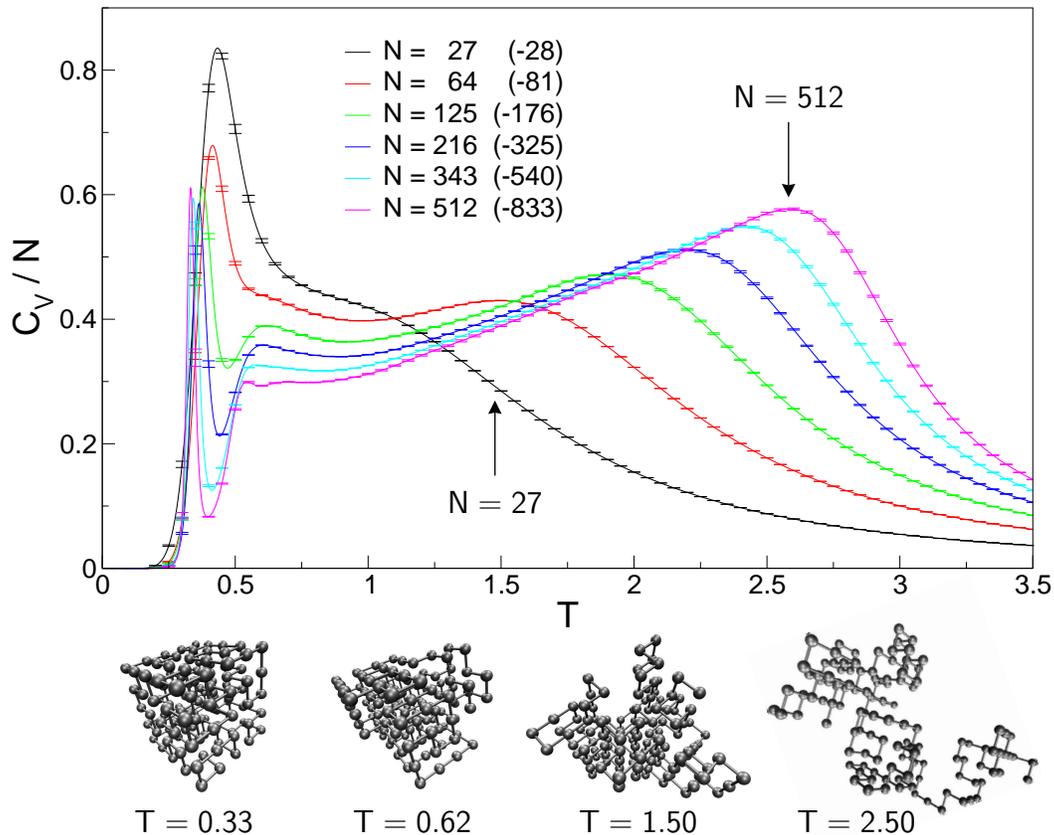}
\caption{\label{figure1}(Color online) Specific heat as a function of
  temperature for lattice homopolymers of various chain lengths $N$ on
  the simple cubic lattice. \textbf{Bottom rows:} Representative
  structures at specific temperatures for $N=125$. (From W\"ust and
  Landau \cite{wuest})}
\end{figure}

Since all monomers are equivalent, obtaining results for multiple
values of the monomer number $N$ will enable us to attempt to
determine the systematic behavior of the polymer collapse as the
length diverges. However, the length dependence is not at all simple.
Those homopolymers of length $N=L^3$, where $L$ is an integer, are
special in that their ground state is a perfect cube of monomers.
(Note, because of the constraints due to the bond connectivity, the
attainment of this cubic overall structure involves quite complicated
configurations.) If the polymer is slightly longer or shorter than a
``magic number'', the behavior may change dramatically because of the
extra degrees of freedom for those monomers that lie outside a perfect
cube. For now, we will concentrate on homopolymers whose length is a
``magic number'' so that we can compare behaviors as the length
increases. The specific heat demonstrates the complexity of the
behavior as the temperature is reduced. First, as shown in
Fig.~\ref{figure1} for $N=27$ a high temperature shoulder with a low
temperature pronounced peak is found. For $N=64$ a rounded maximum
appears to signal the polymer collapse into a globule. At much lower
temperature, however, another, sharper peak announces the
rearrangement of bonds to allow the polymer to assume a configuration
that is very close to the ground state, i.e. there is kind of a
``freezing transition''. With increasing chain length the upper peak 
grows in magnitude monotonically and shifts, systematically to higher 
temperature. In contrast, as the chain length increases, the lower 
temperature peak shifts downwards in temperature and decreases in 
magnitude. This shows clearly that in the present model (simple cubic 
lattice homopolymer model) no symmetry breaking is taking place 
between a dense liquid and a crystal state (both have simple cubic 
symmetry). The origin and sharpness of the low temperature peak is 
rather the result of geometrical surface reconstructions which involve 
the simultaneous displacement of many monomers.  To provide some 
insight into the homopolymer behavior, in the lower part of the figure 
we show ``snapshots'' of typical states of an $N=125$ homopolymer at 
different temperatures above and below the two peaks. In addition, 
there is interesting structure developing just above this sharp, low
temperature peak. At first a small, broadened peak appears, but it
then flattens out and the specific heat then begins to show small
oscillations which appear to be larger than the statistical error
bars. According to ``snapshots'' of the homopolymer structure, these
small oscillations seem to result from shifts of compact portions of
the ground state cube to create distorted surfaces. Clearly, high
resolution results for still longer chains would be highly desirable
to illuminate the actual behavior at, and just above, the low temperature transition. (To be published.)

\section{Structural and thermodynamic properties of some HP model proteins}

\subsection{Free HP proteins}

Unlike the simple, lattice homopolymers, the monomers of HP lattice
proteins are not all equivalent. Each protein has a specific sequence
of $H$'s and $P$'s, and there is no way to extrapolate the behavior to
that of an infinite length equivalent. A number of specific choices
of length and sequence have served as ``benchmarks'' in the
literature; and, for pedagogical reasons, we shall emphasize only a
few three dimensional systems in the current treatment. The sequence
3D103
(P\-P\-H\-H\-P\-P\-P\-P\-P\-H\-H\-P\-P\-H\-H\-P\-H\-P\-P\-H\-P\-P\-P\-P\-P\-P\-P\-H\-P\-P\-P\-H\-H\-P\-H\-H\-P\-P\-P\-P\-P\-P\-H\-P\-P\-H\-P\-H\-P\-P\-H\-P\-P\-P\-P\-P\-H\-H\-H\-P\-P\-P\-P\-H\-H\-P\-H\-H\-P\-P\-P\-P\-P\-H\-H\-P\-P\-P\-P\-H\-H\-H\-H\-P\-H\-P\-P\-P\-P\-P\-P\-P\-P\-H\-H\-H\-H\-H\-P\-P\-H\-P\-P)
is a benchmark sequence whose native state has proven to be
particularly difficult to determine, so we will now concentrate our
discussion on this system. (We note, however, that a great deal of
work has also been carried out on two dimensional sequences as well.)

For very short HP sequences, the native state is accessible, in spite
of the very large number of distinct configurations that result
because of the bond connectivity constraints. For some protein models
the native state is degenerate, and in some cases there are multiple
sequences of the same length that give different results for both
native state energies and temperature dependent behavior. For
example, Bachmann and Janke~\cite{mccg} examined several different
14mer HP sequences by exact enumeration and found noticeably different
specific heat curves. For longer sequences, however, the problem has
proven to be much more challenging. As an example of the relative
success that various computational methods have had in finding native
state energies for current benchmark HP sequences, in Table~\ref{Seqs}
we show results from diverse numerical methods for the lowest energies
found. The table includes results not only for 3D103 but also for
benchmark sequences in two dimensions with almost the same number of
monomers. We note that 2D100 actually represents two different
benchmarks with the same number of monomers but different sequences.
Furthermore, the ground state energies of 2D100a and 2D100b are
\textbf{not} the same. In addition to values from Wang-Landau
sampling (WLS), we also show results from several other sophisticated
approaches: equi-energy sampler (EES)~\cite{ees}, multicanonical chain
growth(MCCG)~\cite{mccg,prellberg}, multi-overlap ensemble
(MSOE)~\cite{msoe}, fragment-regrowth Monte Carlo (FRESS)
\cite{fress}, and pruned-enriched Rosenbluth sampling (PERM)
\cite{perm}. Note that for ``fairness'' reasons only methods which are
capable (at least principally, see \cite{fress}) to sample
thermodynamic quantities or even the entire DOS are included in this
comparison, whereas methods solely targeting the search for the
optimal ground state are excluded (such as e.g. the very efficient
constraint-based approach for protein structure prediction
\cite{backofen}).

\begin{table}
\caption{\label{Seqs}Minimum energies found by several methods for
  benchmark HP sequences in 2D and 3D. The first column names the
  sequence (dimension and length), see \cite{fress}. For
  abbreviations, see text.}
\begin{minipage}{\textwidth}
\begin{tabular}{lcccccc}
\hline\noalign{\smallskip}
Seq. & WLS & EES & MCCG & MSOE & FRESS\footnote{Ground state search only.} & PERM\\
\noalign{\smallskip}\hline\noalign{\smallskip}
2D100a & -48 & -48 &  --  & -47  &  -48  & -48 \\
2D100b & -50 & -49 &  --  & -50\footnote{DOS not attained.}  &  -50  & -50 \\
3D88   & -72 &  -- &  --  &  --  &  -72  & -69 \\
3D103  & -58 &  -- & -56  &  --  &  -57  & -55 \\
3D124  & -75 &  -- &  --  &  --  &  -75  & -71 \\
3D136  & -83 &  -- &  --  &  --  &  -83  & -80 \\
\noalign{\smallskip}\hline
\end{tabular}
\end{minipage}
\end{table}

\begin{figure}
\centering
\includegraphics[width=0.85\columnwidth]{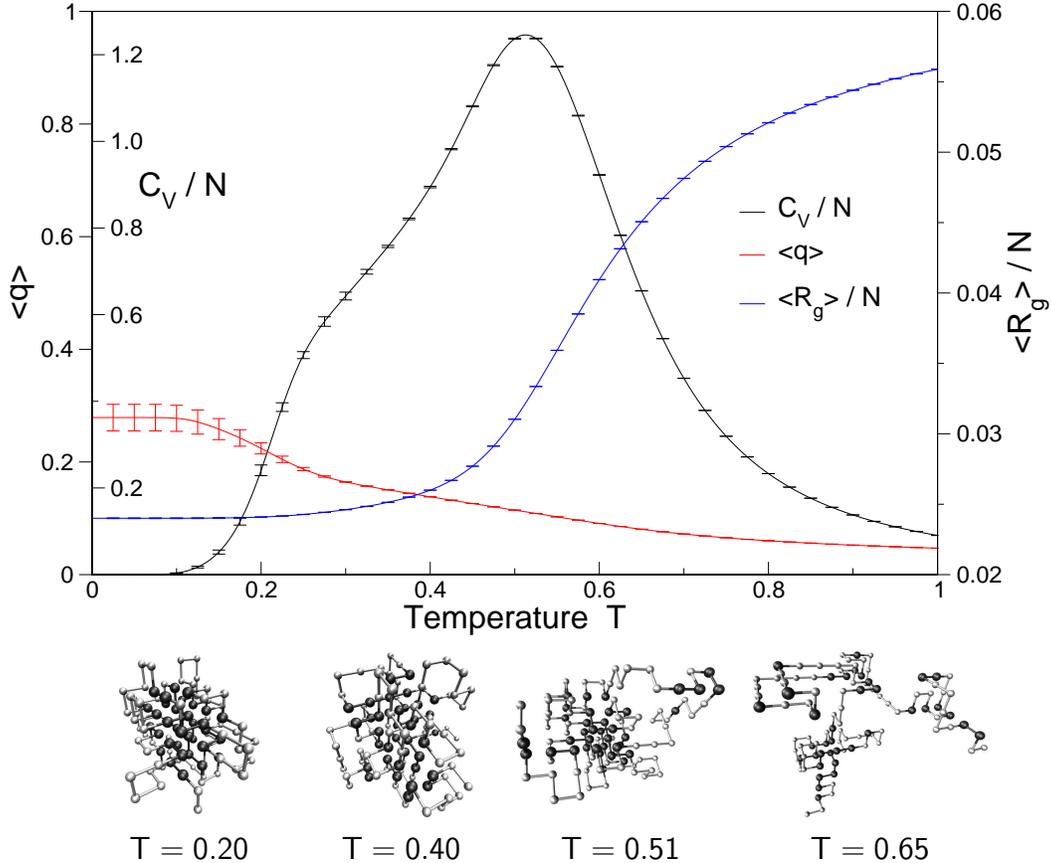}
\caption{\label{figure2}(Color online) Specific heat $C_V/N$, mean
  radius of gyration $\langle R_g\rangle/N$ ($N$, chain length), and
  mean Jaccard index $\langle q\rangle$ as a function of temperature
  for HP sequence 3D103. (From W\"ust and Landau \cite{wuest})
  \textbf{Bottom rows:} Typical conformations at specific
  temperatures. The dark spheres represent $H$-monomers and the light
  spheres are the $P$-monomers.}
\end{figure}

The characteristic behavior of the specific heat as determined by
Wang-Landau sampling is shown for sequence 3D103 in Fig.
\ref{figure2}. At high temperature the protein is distended, but as
the temperature is lowered there is a rounded peak that signals the
collapse of the protein into a compact object, but the acquisition of
the native state only begins at a lower temperature where a shoulder
occurs in the specific heat. The radius of gyration, also plotted in
Fig. \ref{figure2}, decreases substantially in the vicinity of the
specific heat peak, and sample configurations (shown in the bottom
portion of the figure) confirm both the protein collapse as well as
the lack of a native state until below the temperature at which the
shoulder appears. Attempts to determine the native state energy have
an extensive history and estimates were slowly descending to lower
values with time. The lowest energy obtained so far is $E_{min}=-58$
which was obtained from Wang-Landau sampling, but in spite of the
sophistication of the method, it has not been possible to get a
reliable estimate of the relative probability of the native state and
first excited state.

This native state has a well formed hydrophobic core but no obvious
symmetry to distinguish it from the first excited state. The Jaccard
index measures the overlap between the configuration of an HP chain
and one of the native states \cite{fraser:07}. For 3D103 the Jaccard
index increases as the temperature is lowered, but it saturates at $T
\approx 0.1$ (see Fig.~\ref{figure2}) reflecting the degeneracy of the
native state.

\begin{figure}
\centering
\includegraphics[width=0.95\columnwidth]{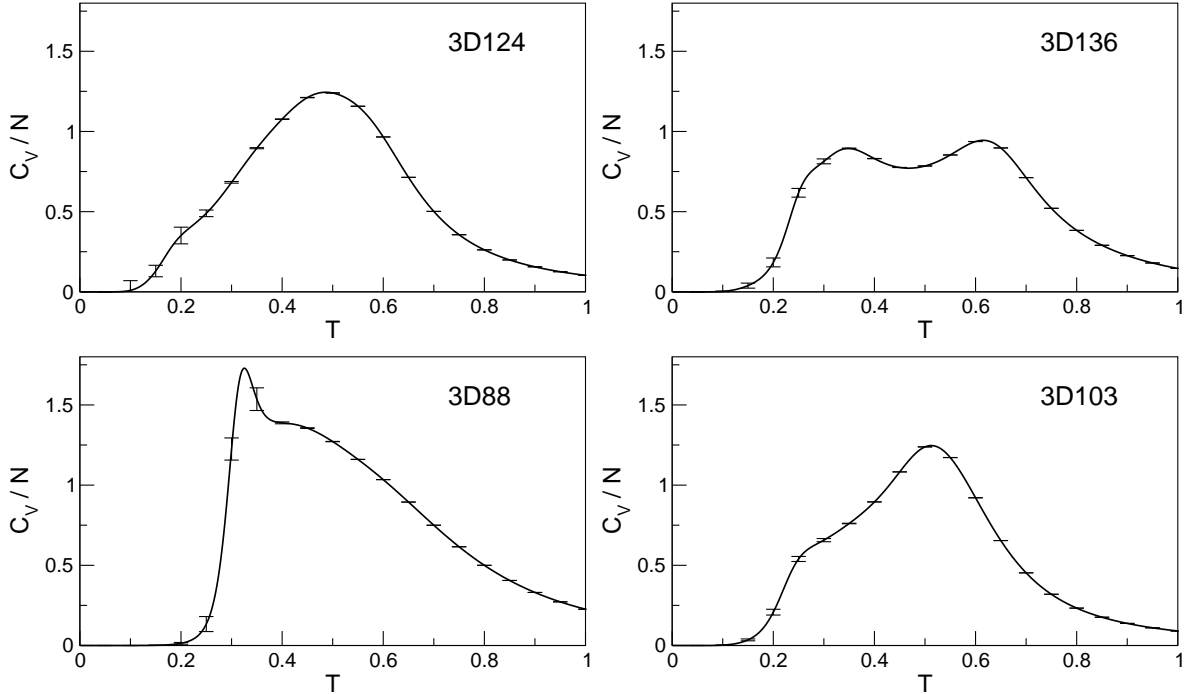}
\caption{\label{hpbench}Specific heat $C_V/N$ as a function of
  temperature ($T$) for the three-dimensional benchmark HP sequences
  listed in Table~\ref{Seqs}. Due to the difficulty to sample the low
  energies, for sequences 3D103 and 3D136, the DOS has been attained
  only to energy levels $-57$ and $-81$, respectively. Error bars have
  been obtained by means of bootstrap resampling.}
\end{figure}

Fig.~\ref{hpbench} shows the specific heat as a function of
temperature for the three-dimensional benchmark HP sequences listed in
Table~\ref{Seqs}. Comparison among these curves shows clearly their
common underlying thermodynamic behavior; that is, the occurrence of a
protein collapse ($C_V$ peak at higher temperatures) and another
``folding transition'' to the native state(s) at low temperatures.
However, the characteristics of the later transition shows significant
sequence and length dependences ranging from a sharp peak (3D88) to a
soft shoulder (3D124) only. This is particularly remarkable in
consideration that for all four sequences the sampling of low
temperature range is equally difficult.

\subsection{Thermodynamic properties of the HP model near an attractive surface}

To demonstrate the richness of the ``transition'' behaviors of the HP
chain in the vicinity of an attractive surface, we examined an HP
sequence with 36 beads (PPPHHPPHHPPPPPHHHHHHHPPHHPPPPHHPPHPP) and a
surface which attracts both $H$- and $P$-monomers. We focus on two
cases: a weak attractive surface ($\varepsilon_{HH} = 12,
\varepsilon_s = 1$) and a strong attractive surface ($\varepsilon_{HH}
= 1, \varepsilon_s = 2$), to compare the effects of different surface
strengths on the ``transition'' behaviors. Technical details have been
discussed in Ref. \cite{li2011}. Similar studies can be found in Refs.
\cite{bachmann_janke_2006,swetnam}.

\begin{figure}
\centering
\includegraphics[width=0.65\columnwidth]{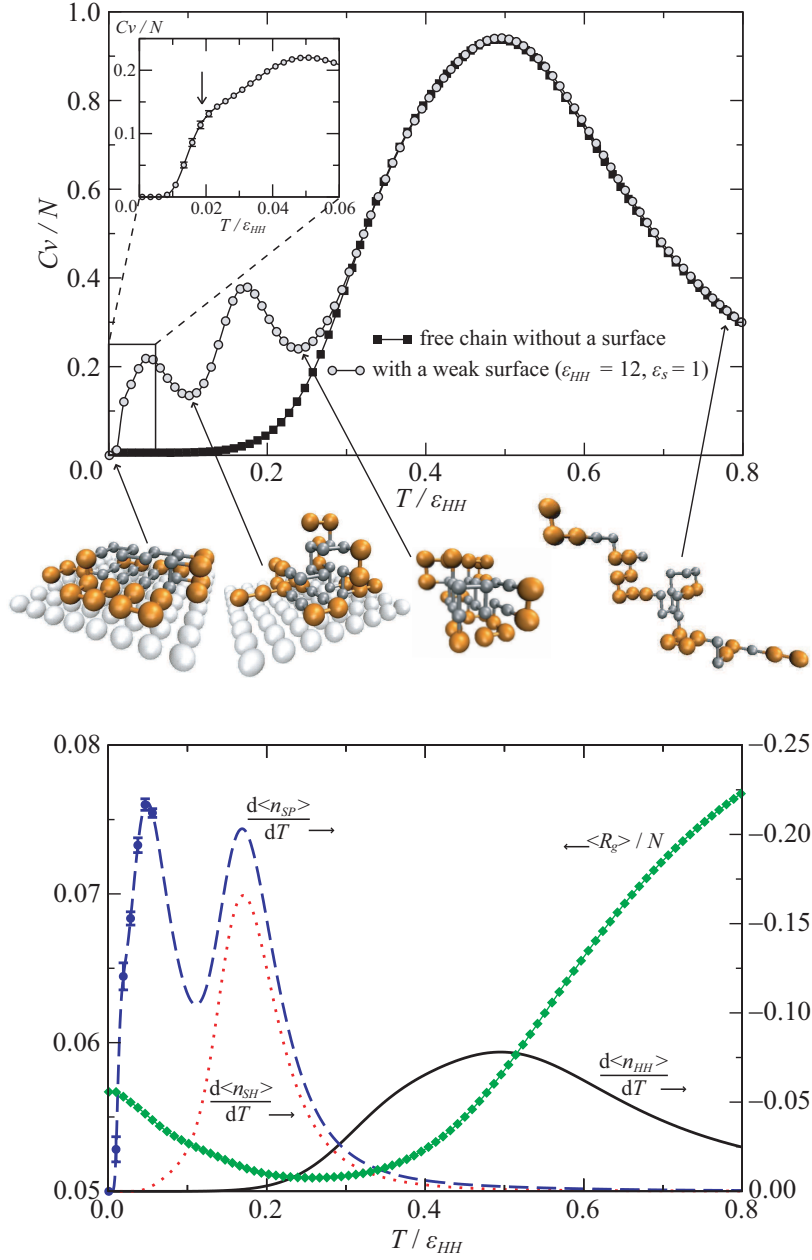}
\caption{(Color online) \textbf{Top:} Specific heat of HP sequence
  36mer interacting with a weak attractive surface ($\varepsilon_{HH}
  = 12, \varepsilon_s = 1$) and without the presence of the
  surface. Typical configurations are shown for several different
  temperatures. The small spheres represent $H$-monomers and the large
  spheres are $P$-monomers. The faint spheres at the bottom for the
  two lowest temperatures show the surface. (From Li, W\"ust and
  Landau \cite{li2011}) \textbf{Bottom:} Radius of gyration and
  thermal derivatives of the numbers of $H$-$H$ contacts as well as
  surface contacts. Horizontal arrows besides the legends indicate the
  scales that the quantities are using. For both graphs, error bars
  smaller than the data points are not shown.}
\label{36mer}
\end{figure}

As in the case of free proteins, the use of ``pull moves'' and ``bond
rebridging'' turn out to be essential for accessing low energy states
relatively rapidly. The top portion of Fig. \ref{36mer} shows a
typical specific heat of an HP chain interacting with a weakly
attractive surface, as compared to that of the free protein sequence
without the surface. In the former case, three peaks are observed to
signal three ``transition'' stages: the largest peak at $T /
\varepsilon_{HH} \approx 0.5$ represents the same coil-globule
transition as found in free space. (The comparison of the specific
heat from the two cases in Fig. \ref{36mer} demonstrates this very
clearly.) During this stage, the HP polymer transforms from an
extended chain-like structure to a compact, but desorbed, globule.
Typical structures are much the same as the collapsed states of the
free chain in the absence of the surface.

The middle peak at $T / \varepsilon_{HH} \approx 0.18$ signals protein
adsorption during which the compact HP globule ``docks'' at the
surface with the hydrophobic core remaining intact. The globule spans
several layers vertically and the total energy of the system is
lowered slightly due to contact with the surface.

Further decrease in temperature brings the system to the third
transition at $T / \varepsilon_{HH} \approx 0.05$ where the system
maximizes the number of surface interactions without sacrificing an
intact, energetically minimized hydrophobic core. Forming $H$-$H$
contacts is energetically much more favorable than forming surface
contacts with a large value of $\varepsilon_{HH}$. The last
interesting feature in the specific heat is a subtle shoulder at $T /
\varepsilon_{HH} \approx 0.02$. Typical states at $T /
\varepsilon_{HH} = 0$ and $T / \varepsilon_{HH} \approx 0.02$ suggests
this is due to the difference in shapes of hydrophobic cores and the
number of $P$-monomer-surface contacts available at these
temperatures. At $T / \varepsilon_{HH} = 0$ where there are just
ground states, only rectangular $H$-cores allow a maximum number of
$P$-surface contacts that minimizes the energy.

The temperature dependence of the structural properties is shown in
the bottom portion of Fig. \ref{36mer}. At the temperature for which
the specific heat shows the largest peak, the radius of gyration is
rapidly decreasing and the number of $H$-$H$ contacts is rapidly
increasing (which results in a peak in its thermal derivative) on
cooling. Both behaviors support the idea of a ``coil-globule''
transformation. At the temperature where the middle specific heat peak
occurs, the thermal derivative of the number of $H$-surface contacts
also shows a peak. The thermal derivative of the number of $P$-surface
contacts has maxima at the same temperatures as both of the low
temperature specific heat peaks. It also shows a weak shoulder at the
same temperature as the very low temperature specific heat shoulder
($T / \varepsilon_{HH} \approx 0.02$). The radius of gyration also
increases at low temperature because the protein is flattening out on
the surface and becoming rather two dimensional in shape.

\begin{figure}
\centering
\includegraphics[width=0.65\columnwidth]{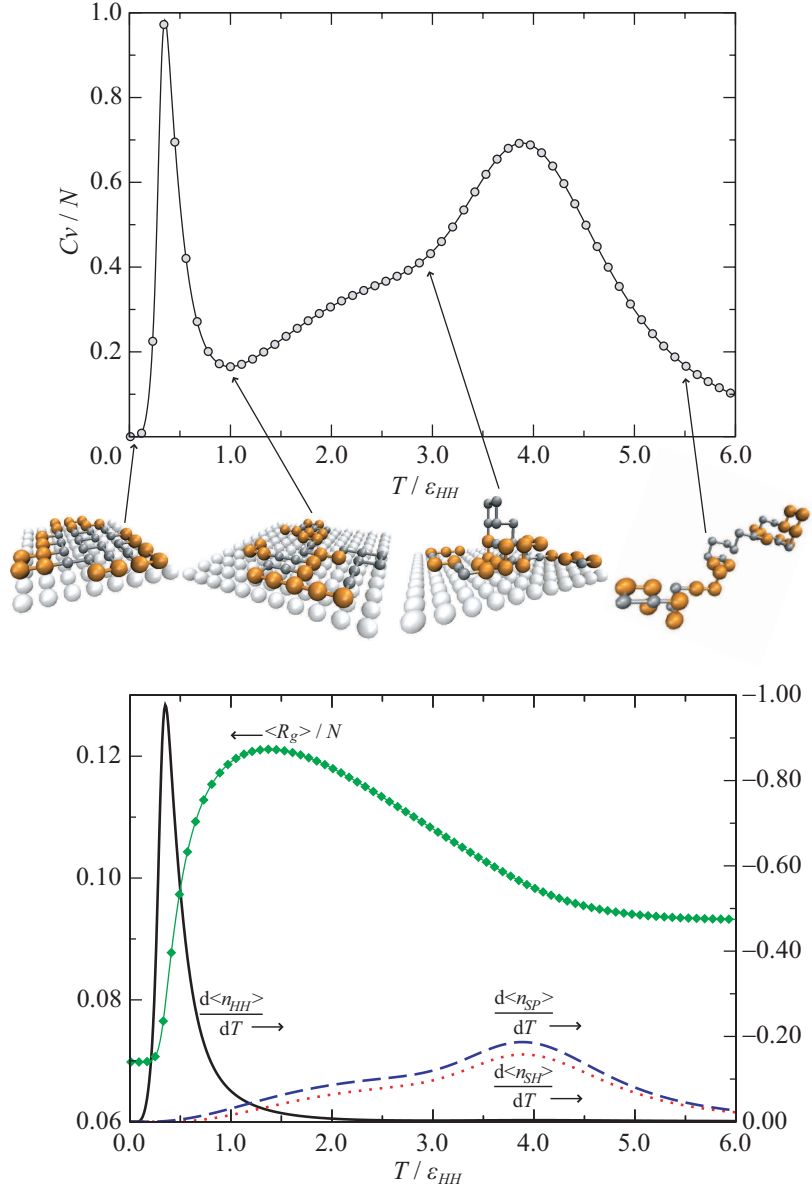}
\caption{(Color online) \textbf{Top:} Specific heat of HP sequence
  36mer interacting with a weak attractive surface ($\varepsilon_{HH}
  = 1, \varepsilon_s = 2$). Typical configurations are shown for several different
  temperatures. The small spheres represent $H$-monomers and the large
  spheres are $P$-monomers. The faint spheres at the bottom for the
  two lowest temperatures show the surface. \textbf{Bottom:} Radius of
  gyration and thermal derivatives of the numbers of $H$-$H$ contacts
  as well as surface contacts. Horizontal arrows besides the legends
  indicate the scales that the quantities are using. For both graphs,
  error bars are not shown as all are smaller than the data points.}
\label{36mer_strong}
\end{figure}

However, if the surface is strongly attractive, the transition
behavior is quite different from the previous case as shown in Fig.
\ref{36mer_strong}. For the surface strength we have chosen, only two
peaks with a weak bump in between are seen in the specific heat. The
peak at $T / \varepsilon_{HH} \approx 3.9$ indicates a protein
adsorption transition, and the peak at $T / \varepsilon_{HH} \approx
0.35$ corresponds to the protein collapse transition that takes place
completely on the surface. The bump at $T / \varepsilon_{HH} \approx
2.0$ then signals the flattening stage.

Understanding this in conjunction with the structural properties, it
is obvious to see that at high temperature where the specific heat
shows a peak, the extended HP chain first adsorbs on the surface. This
is in agreement with the peaks in the derivatives of the number of
surface contacts at the same temperature. In the temperature range $T
/ \varepsilon_{HH} \approx 3$ down to $T / \varepsilon_{HH} \approx
1$, the partly adsorbed chain flattens itself out until all monomers
come in contact with the surface. Again, the specific heat and the
derivatives of the number of surface contacts echo each other by
revealing convex bumps at the same temperature. The radius of gyration
increases rapidly in this temperature range as the structure becomes
more and more planar. However, the chain remains extended without
forming many $H$-$H$ contacts during this stage as no signal is found
from the derivative of the number of $H$-$H$ contacts. As a third step
the fully adsorbed, two-dimensional chain undergoes a ``collapse''
transition at $T / \varepsilon_{HH} \approx 0.35$ to maximize the
number of $H$-$H$ contacts before adopting a film-like, compact
structure. This is clearly signaled by the sharp peaks in the specific
heat and the derivative of the number of $H$-$H$ contacts at that
temperature.

As shown in the two examples here, structural transitions may occur in
different orders for different surface attractive strengths when
temperature changes. Some transitions may not give distinct signals in
specific heat but merely in particular structural parameters. It is
thus essential to analyze specific heat together with structural
quantities to identify various ``phases''. With longer HP chains, our
interpretation of selected structural parameters suggests that the
hierarchies of structural phase transitions for this model can be
generalized into a few categories. (To be
published.)

\section{Discussion and Conclusion}

Homopolymer collapse and protein folding remain challenging problems,
even for the prototypical lattice models discussed in this article.
We described the implementation of an efficient Monte Carlo algorithm
(Wang-Landau sampling) to calculate the density of states directly for
both homopolymers and HP models with different lengths and monomer
sequences. In both cases there were a combination of energetic and
entropic barriers that made the convergence difficult. These problems
could be largely overcome with Wang-Landau sampling with a combination
of pull moves and bond-rebridging moves and with the choice of a
relatively stringent final modification factor. Examination of a
combination of thermodynamic and structural properties, along with
visualization of typical configurations, could be interpreted
coherently to explain the physical behavior that was taking place,
both with and without the presence of a surface. Other HP sequences 
remain to be studied, and the low temperature behavior of even longer 
homopolymers requires careful examination.

\begin{acknowledgments}

We thank M. Bachmann and W. Janke for illuminating discussions. The
research project is supported by the National Science Foundation under
Grant No. DMR-0810223.

\end{acknowledgments}




\end{document}